\documentclass{ws-procs961x669}            % default, citations in superscript
\begin{document}
\title{Tree-Unitarity and renormalizability in Lifshitz-scaling theory\\ 
- as a toy model of Ho\v{r}ava's gravity theory -}

\author{Toshiaki Fujimori}

\address{Department of Physics, and Research and Education Center for Natural
 Sciences, \\ Keio University, Hiyoshi 4-4-1, Kanagawa 223-8521, Japan}

\author{Takeo Inami}

\address{Department of Physics, National Taiwan University, Taipei 10617, Taiwan, R.O.C.\\
Theoretical Research Division, Nishina Center, RIKEN, Wako 351-0198, Japan}

\author{Keisuke Izumi}

\address{Departament de F{\'\i}sica Fonamental, Institut de Ci\`encies del Cosmos (ICCUB),
\\
Universitat de Barcelona, Mart\'{\i} i Franqu\`es 1, E-08028 Barcelona, Spain}

\author{Tomotaka Kitamura}

\address{Department of physics, Waseda University, Shinjyuku, Tokyo 169-8555, Japan}

\begin{abstract}
We study tree-unitarity and renormalizability in Lifshitz-scaling theory, 
which is characterized by an anisotropic scaling between the spacial and time directions.
Due to the lack of the Lorentz symmetry, the conditions for both unitarity and renormalizability are modified from those in relativistic theories. 
For renormalizability, the conventional discussion of the power counting conditions has to be extended. 
Because of the dependence of $S$-matrix elements on the reference frame, 
unitarity requires stronger conditions than those in relativistic cases. 
We show that the conditions for unitarity and renormalizabilty are identical as in relativistic theories.  
We discuss the importance of symmetries for a theory to be renormalizable.
\end{abstract}


\bodymatter

\section{Introduction}\label{aba:sec1}

A quantum field theory requires both unitarity and renormalizability to be a well-defined theory. 
%These properties could fix the form of consistent interactions. 
They may be implied from each other, which motivated the study of the equivalence between unitarity and renormalizability in gauge theories~\cite{Llewellyn Smith:1973ey} and gravitational theories~\cite{Berends:1974gk}. 
This relation could be helpful to construct a consistent quantum gravity; 
to check the renormalizability  we investigate unitarity and vice versa.
It is worth checking whether the relation is a generic property of quantum theories. 
To avoid the complication due to symmetries, it is better to investigate theories with few symmetries. 
One of ways to loosen a symmetry is considering non-relativistic theory. 

Lifshitz-type theories~\cite{Lifshitz scalar} are characterized by the Lifshitz scaling:
\begin{eqnarray}
t \to b^z t,  \hspace{2cm} x^i \to bx^i \hspace{5mm} (i=1,\dots d), \label{1}
\end{eqnarray}
and manifestly do not have Lorentz symmetry. 
These theories have a lot of attention in cosmology~\cite{Izumi:2010yn} and some of quantum properties are also investigated~\cite{Fujimori:2015wda,Fujimori:2015mea,Anselmi:2007ri,Visser:2009fg,Colombo:2014lta}.
Due to the anisotropic scaling (\ref{1}), the ultraviolet(UV) behaviors are different from those in relativistic theories. 
This idea has been introduced to gravitational theory, which is called Ho\v{r}ava-Lifshitz (HL) gravity~\cite{Horava:2009uw}. 
The application of the HL gravity is widely studied in cosmology (see Ref.~\citenum{Mukohyama:2010xz} for a review), 
such as the emergence of dark matter as an integration constant \cite{DM} and so on.
Its quantum properties have not been studied very much~\cite{Barvinsky:2015kil}.
By the power-counting discussion HL gravity is expected to be renormalizable, but the renormalizability still remains uncertain. 

We have studied tree-unitarity and renormalizability in theories of a Lifshitz scalar field to investigate the relation between them~\cite{Fujimori:2015wda,Fujimori:2015mea}.
These thoeries are also useful as toy models of HL gravity theory. 
The lack of the Lorentz symmetry gives rise to significant modifications to both renormalizability and unitarity.
We have shown that, for theories to be renormalizable, the conditions for renormalization have to be extended from the conventional ones. 
As for tree-level unitarity, since scattering amplitudes have the reference frame dependence, unlike in relativistic cases, 
stronger conditions are required. 
We have derived  the extended power-counting renomalizability (PRC) conditions and the tree-level unitarity constraints 
on the forms of self-interaction term in the theory of a single Lifshitz scalar field. 
We have seen the equivalence between these conditions in cubic and quartic interactions. 
 
We give a brief review of our papers Refs.~\citenum{Fujimori:2015wda,Fujimori:2015mea}. For details, see these papers. We also discuss a consequence of our argument to the renormalizability of HL gravity.

\section{Lifshitz scalar field}\label{aba:sec2}

The quadratic action for Lifshitz scalar field $\phi$ can be written in the form
\begin{eqnarray}
S_2 = \int dt d^d x \bigg[ \frac{1}{2}\phi \Big\{ \partial_t^2 - f(-\triangle) \Big\} \phi  \bigg], 
\qquad \Delta:= \partial^i\partial_i, 
\label{2nd}
\end{eqnarray}
where $f(-\triangle)= (-\triangle)^z +\alpha(-\triangle)^{z-1} + \dots$ is a polynomial of degree $z$.  
The dispersion relation takes the form
\begin{eqnarray}
E = \sqrt{f(p^2)},
\label{DR}
\end{eqnarray}
where $E$ and $p$ are the energy and the magnitude of the momentum, respectively.
This dispersion relation becomes $E \approx p^z$ in the UV regime
and is consistent with the Lifshitz scaling (\ref{1}). 
Since the asymptotic behavior of propagators is given by 
\begin{eqnarray}
\frac{1}{E^2 - f(p^2)} \approx \frac{1}{E^2-p^{2z}},
\end{eqnarray}
UV divergences of loop diagrams are milder 
than those in relativistic theories for $z>1$. 

Let us introduce the following generic $n$-point interaction term
\begin{eqnarray}
S_{int} = \lambda \int dt d^d x \, (\partial_x^{a_1} \phi) \, (\partial_x^{a_2} \phi) \cdots (\partial_x^{a_n} \phi), 
\label{eq:vertex}
\end{eqnarray}
where $a_1,\cdots, a_n$ are non-negative integers. 
Each $\partial_x$ denotes any of the spatial derivatives $\partial_i~(i=1,\cdots,d)$ 
and we suppose that this interaction term is designed 
to keep the $d$-dimensional spatially rotational symmetry $O(d)$.

\section{Renormalizability}\label{aba:sec3}

The scaling dimension of $\phi$ can be read from the quadratic term in (\ref{2nd}):
\begin{eqnarray}
[\phi] = \frac{d-z}{2},
\label{eqphi}
\end{eqnarray}
where we have used the convention that $[E]=z$ and $[p]=1$. 
The dimension of the coupling constant $\lambda$ is obtained as
\begin{eqnarray}
[\lambda] = z + d - \sum_{l=1}^n ( a_l + [\phi] ). 
\label{eqlam}
\end{eqnarray}
In relativistic theories, 
an interaction term is renormalizable if it has a coupling constant with non-negative dimension: $[\lambda] \geq 0$.
On the other hand, in the Lifshitz scalar theory, 
non-negativity of $[\lambda]$ is not enough for renormalizability.

As an example, let us consider the case with $d=3$ and $z=5$, and the following quartic  interaction term with $[\lambda]=0$, 
\begin{eqnarray}
S_4 = \lambda \int dt d^3x \, \phi^2 ( \Delta^3 \phi )^2.
\label{4th}
\end{eqnarray}
Let us consider an 1-loop diagram with $n$ vertices  each of which has two external lines. 
The term for which two $( \Delta^3 \phi )$'s at each vertex correspond to the internal lines gives the leading order contribution.
This can be estimated as 
\begin{eqnarray}
\int d\omega d^3 p \left(\frac{1}{\omega^2-p^{10}}\right)^n \left( p^{12}\right)^n
\sim \Lambda^{8+2n},
\end{eqnarray}
where $\Lambda$ is a UV cutoff with $[\Lambda]=1$. 
For any $n$, this loop integral diverges as $\Lambda \to \infty$, 
and thus the interaction term \eqref{4th} is NOT renormalizable, even though the condition $[\lambda] \geq 0$ is satisfied. 

The reason why the interaction term \eqref{4th} is not renormalizable is as follows. 
In the calculation of loop integrals, only operators corresponding to the internal lines are involved. 
Thus, only their dimension is related to the renormalizability. 
In the above example, the dimension of field is negative ($[\phi]=-1$) and the part of the term $( \Delta^3 \phi )^2$ has smaller dimension than the whole of the term $\phi^2 ( \Delta^3 \phi )^2$. 
Now, the whole of $\phi^2 ( \Delta^3 \phi )^2$ is marginal, and thus the part of the term $( \Delta^3 \phi )^2$ is non-renormalizable.
Therefor, the criteria of renormalizability is; the dimension of any part of the term has to be larger than the inverse of $[dt d^d x]$.\footnote{
For whole of the term, the case of equality can also be renormalizable. This is the conventional PCR.
}
For the proof in generic cases, see Ref.~\citenum{Fujimori:2015mea}.

\section{Tree-unitarity}\label{aba:sec4}

The unitarity bound is one of the necessary conditions for the untarity of $S$-matrix elements. 
The unitarity condition reduces to the optical theorem
\begin{eqnarray}
2{\rm{Im}}{\cal{M}}\left(i\to i\right)=\sum_{X}\delta(E-E_X) \delta^{d} \left( {\bf p} - {\bf p}_X \right)|{\cal{M}}\left(i\to X\right)|^2,
\label{op}
\end{eqnarray}
where ${\cal{M}}\left(i\to j\right)$ is the scattering amplitude, $\{ |{X} \rangle \}$ be a complete orthonormal basis of the Hilbert space:
\begin{eqnarray}
\sum_{X}|X\rangle \langle X|=1,
\end{eqnarray}
$i$ is an element in $\{ |{X} \rangle \}$, and $\sum_{X}$ denotes the sum with respect to all possible intermediate states. 
In the following, we ignore unimportant numerical factors.
Taking absolute value of eq.(\ref{op}), the left hand side has to be smaller than or equal to $\left|{\cal{M}}\left(i\to i\right)\right|$. 
Since the right hand side is the summation of non-negative values, each of them has to be smaller than or equal to the sum. 
Then, we have the unitarity bound:
\begin{eqnarray}
\left|{\cal{M}}\left(i\to i\right)\right|\ge\delta(E-E_X) \delta^{d} \left( {\bf p} - {\bf p}_X \right)|{\cal{M}}\left(i\to j\right)|^2,
\label{uni}
\end{eqnarray}
where $j$ is an element in $\{ |{X} \rangle \}$ and can be the same as $i$. 

The forms of cubic and quartic interaction terms can be constrained with the uniarity bound of two-particle scattering. 
We take the basis labeled with the total energy and momentum;
\begin{eqnarray}
|E,{\bf P},l\rangle = \int \frac{d^dp_{2}}{2E_{2}}\frac{d^dp_{2}}{2E_{2}}\delta \left({E_1}+{E_2}-E\right)\delta^d\left({\bf{p}}_1+{{\bf{p}}_2}-{\bf{P}}\right) \, h_l\left( {\bf p}_j \right)|{{\bf p}}_1,{\bf{p}}_2\rangle, 
\label{epl}
\end{eqnarray}
where $|{{\bf p}}_1,{\bf{p}}_2\rangle$ is the standard 
asymptotic momentum eigenstates for two-particle states normalized as
\begin{eqnarray}
\int \frac{d^{d} p_1}{2E_1}\frac{d^{d} p_2}{2E_2}|{\bf{p}}_1, \cdots,{\bf{p}}_n \rangle \langle{\bf{p}}_1, \cdots,{\bf{p}}_n|=1,
\end{eqnarray}
in two-particle basis of subspace, 
$E_1$ and $E_2$ are the energy of particles, 
and $\{h_l\left({\bf p}_j \right)\}$ is an orthonormal functions in two-particle subspace with the total energy $E$ and momentum ${\bf P}$. 
In the two-particle subspace, the unitarity bound (\ref{uni}) can be rewritten as 
\begin{eqnarray}
|{\cal{M}}\left(E,{\bf{P}};{l} \to {l'}\right)| \leq \text{const}.
\label{UB}
\end{eqnarray}
This is a necessary condition for the unitarity.
We check the high-energy behavior of this.

In relativistic theories, we can always take the center-of-mass frame, i.e. ${\bf{P}}={\bf{0}}$. 
Then, both particles have the same magnitude of momentum. 
In the high energy limit, the energies of both particles go to infinity. 
In the theories without Lorentz symmetry, in contrast, 
states with nonzero total momentum has no relation to those with ${\bf{P}}={\bf{0}}$. 
We should also consider the situation where only the energy of one particle goes to infinity and the other is finite. 
Moreover, this state can transfer to the state where energies of both particle are infinity. 
Because of these varieties, we have more constraints. 
From the unitarity bound (\ref{UB}) with the tree-level approximation, we have the same constraints with the extended version of PCR.
We omit the derivation of the constraints here. For details, see  Ref.~\citenum{Fujimori:2015mea}.

\section{Discussion -importance of symmetry-}\label{aba:sec5}

We have seen that, if the dimension of fields is non-positive, the extension of PCR is required. 
This is also true for unitarity, since, as we have demonstrated, the conditions for unitarity are identical to those for renormalizability. 
In Ref.~\citenum{Fujimori:2015wda}, we have addressed the importance of symmetries for a theory to naturally have renormalizability. 
We have studied theories with a dimensionless scalar field and shown that 
in some theories with symmetry, only the conventional conditions of PCR could be enough. 
The conventional PCR shows that, in theories with dimensionless fields $\phi$, 
marginal interaction terms require the counter terms having all relevant or marginal forms. 
Because the product of a marginal term and $\phi^n$ is marginal for any $n$, we have an infinite number of marginal terms. 
However, 
introducing the shift symmetry which is the invariance under $\phi\to \phi+$const., the scalar field $\phi$ must always appear in combination with space-time derivatives and the dimension of $\partial_\mu \phi$ is positive. 
Therefore, all possible operators have positive dimension and PCR does NOT need to be extended. 

Based on the above discussion, HL gravity with $d=z=3$ is expected to be renormalizable. 
If we separately see the interaction terms, HL gravity has terms violating the extended version of PCR. 
However, because of the symmetry, the possible operators are the three-dimensional curvature $R_{ij}$ and the extrinsic curvature $K_{ij}$ with or without differential operators. 
These operators have positive dimension, and the number of relevant and marginal terms generated by loop becomes finite. 
Therefore, the conventional PCR could work well. 

The extended version of PCR is important when we introduce matter fields in HL gravity or consider higher $z$. 
To keep the scaling of gravitational field, matter fields should satisfy the same scaling, and then they must be dimensionless for $d=z=3$. 
Without symmetry of matter sector, the matter fields have to satisfy the extended PCR. 
Imposing symmetries could naturally lead the theory renormalizable as in the above discussion of HL gravity.  
For higher $z$, for instance in the case with $d=3$ and $z\ge7$, the dimension of curvature is non-positive, that is, the symmetry does not restrict the possible operator to be positive. 
In this case, the extended  PCR is required.

\section*{Acknowledgments}

The work of T. F. is supported in part by the Japan Society for the Promotion of Science
(JSPS) Grantin-Aid for Scientific Research (KAKENHI Grant No. 25400268) and by the MEXTSupported
Program for the Strategic Research Foundation at Private Universities ``Topological
Science''(Grant No. S1511006). 
K.~I. is supported by FPA2013-46570-C2-2-P, AGAUR
2009-SGR-168, and MDM-2014-0369 of ICCUB (Unidad de Excelencia `Maria de Maeztu').

\end{document}